\documentclass[11pt]{article}
\usepackage{euscript}
\usepackage{amssymb}
\usepackage{amsfonts}
\usepackage{amsbsy}
\usepackage{epsfig}
\usepackage{amsthm}
\usepackage{amscd}
\usepackage{amstext}
\usepackage{verbatim}
\usepackage{amsmath}
\usepackage{hyperref}
\usepackage{cite}

\pdfoutput=1

\begin{document}

\begin{center}
{\Large \textbf{Holographic RG flow of thermo-electric transports with
momentum dissipation }}

\vspace{1cm}

Shao-Feng Wu$^{1}$, Bin Wang$^{2,3}$, Xian-Hui Ge$^{1,4}$, Yu Tian$^{5,6}$

\vspace{1cm}

{\small \textit{$^{1}$Department of physics, Shanghai University, Shanghai,
200444, China }}\\[0pt]
{\small \textit{$^{2}$Center for Gravitation and Cosmology, Yangzhou
University, Yangzhou 225009, China}}\\[0pt]
{\small \textit{$^{3}$Department of Physics and Astronomy, Shanghai Jiaotong
University, Shanghai, 200240, China}}\\[0pt]
{\small \textit{$^{4}$Department of Physics, University of California at San
Diego, CA92093, USA}}\\[0pt]
{\small \textit{$^{5}$School of Physics, University of Chinese Academy of
Sciences, Beijing, 100049, China}}\\[0pt]
{\small \textit{$^{6}$Institute of Theoretical Physics, Chinese Academy of
Sciences, Beijing, 100190, China}}\\[0pt]
\vspace{0.5cm}

{\small \textit{sfwu@shu.edu.cn, wang\_b@sjtu.edu.cn, gexh@shu.edu.cn,
ytian@ucas.ac.cn}}
\end{center}

\vspace{1cm}

\begin{abstract}
We construct the holographic renormalization group (RG) flow of
thermo-electric conductivities when the translational symmetry is broken.
The RG flow is probed by the intrinsic observers hovering on the sliding
radial membranes. We obtain the RG flow by solving a matrix-form Riccati
equation. The RG flow provides a high-efficient numerical method to
calculate the thermo-electric conductivities of strongly coupled systems
with momentum dissipation. As an illustration, we recover the AC
thermo-electric conductivities in the Einstein-Maxwell-axion model.
Moreover, in several homogeneous and isotropic holographic models which
dissipate the momentum and have the finite density, it is found that the RG
flow of a particular combination of DC thermo-electric conductivities does
not run. As a result, the DC thermal conductivity on the boundary field
theory can be derived analytically, without using the conserved thermal
current.
\end{abstract}
\pagebreak

\section{Introduction}

\textquotedblleft GR=RG\textquotedblright\ \cite{ZaanenBook}. In the
holographic theory, this short \textquotedblleft equation\textquotedblright\
highlights that the renormalization group (RG), an iterative coarse-graining
scheme to extract the relevant physics \cite{Wilson74,Wilson83,Polchinski84}%
, is essential in generating the bulk gravity dual from the boundary field
theory. Although the precise process of coarse graining is not clear, it is
evident that the anti-de Sitter/conformal field theory (AdS/CFT)
correspondence provides the geometrisation of RG flow, in which the radial
direction in the bulk can be identified with certain energy scale \cite%
{Maldacena97,Gubser98,Witten98,Susskind98,Peet98,Akhmedov98,Alvarez98,Girardello98,Distler98,Balasubramanian99,Freedman99,Verlinde99,Boer01}%
. As an important implication of this picture, one can expect that some
low-energy universality of strongly coupled systems is captured by the
near-horizon degrees of freedom alone.

On the other hand, as Pauli said, \textquotedblleft Solid state physics is
dirty\textquotedblright . The disorder is one of the fundamental themes in
condensed matter theories (CMT). It is an important progress that the
AdS/CMT duality can dissipate the momentum and thereby get close to the real
materials. The simplest way to break the translational symmetry in the
holographic theories is to introduce the linear axion fields \cite%
{Andrade1311}.

Recently, some of us studied the holographic RG flow for the strongly
coupled systems with finite density and disorder \cite{Ge1606,Ge16062,Tian17}%
, where the charge and energy transport are coupled and the transport
coefficients are finite. The partial motivation of the work came from Ref.
\cite{Sin1109}, where the authors have not obtained the explicit flow when
facing with the coupled transport. By introducing a square matrix of coupled
sources, we illustrated that the coupled second-order equations of linear
perturbations can be reduced to a first-order matrix Riccati equation, which
can have the direct physical meaning of the RG flow equation of two-point
correlation functions \cite{Tian17}. In addition, the boundary condition of
the matrix Riccati equation can be simply determined by the regularity of
correlation functions on the horizon. As a result, the holographic RG flow
provides a new method for calculating the coupled transport in holographic
systems, particularly with translational symmetry breaking. Compared with
the traditional method that solves the coupled second-order perturbation
equations directly, the new method can greatly simplify the numerical
calculation, particularly for the AC transport or spatially inhomogeneous
systems. This is mainly because it only needs a simple Runge-Kutta marching
instead of the inconvenient shooting method or the resource-consuming
pseudo-spectral method.

In \cite{Ge1606,Ge16062,Tian17}, however, the holographic RG flow was mainly
used to study the DC transport on the boundary. In this paper, the first aim
is to translate the matrix Riccati equation to the RG flow of AC
thermo-electric conductivities, from which one can read the AC
thermo-electric conductivities on the boundary. As an illustration, we will
calculate these conductivities in the Einstein-Maxwell-axion (EMA) model.
The results are in agreement with the previous work \cite{Sin1409} that
solves the coupled second-order equations directly.

The second aim is to explore whether the holographic RG flow could imply
some interesting physics about the thermo-electric transport in strongly
coupled systems. One important lesson learned from the studies on the
holographic RG flow is that the universality of the transport in the
holographic models may be correlated to the similarity of all horizons and
the existence of certain quantities which do not evolve between the horizon
and the boundary \cite{Liu0809}. Two benchmark examples are the trivial RG
flow of the DC electrical conductivity for the systems dual to neutral black
holes and the ratio between shear viscosity and entropy density in a wide
class of holographic theories. Notably, the trivial RG flow interpolates the
classical black hole membrane paradigm \cite{Thorne86,Wilczek97} and AdS/CFT
smoothly. Based on this universality argument, Blake and Tong identified a
massless mode in the massive gravity and obtained the analytical expression
of the DC electric conductivity \cite{Tong1308}. Furthermore, Donos and
Gauntlett constructed the electric and thermal currents that are radially
conserved. Combined with the choice of sources that are linear in time, they
found an analytical relation between the DC thermo-electric conductivities
on the boundary and the black hole horizon data \cite{Donos1406}. However,
unlike the conserved electric current that usually can be read from the
Maxwell equation, the construction of the conserved thermal current is
considerably more subtle. Noticing this problem, Liu, Lu, and Pope recently
suspected that the Noether current with respect to the diffeomorphism
symmetry might be a general formula for the radially conserved thermal
current \cite{Lu1708}.

We will show that the RG flow of a particular combination of DC
thermo-electric conductivities, namely, the electrical conductivity at zero
heat current, does not run in several homogeneous and isotropic holographic
models which dissipate the momentum and have the finite density. Since the
zero-heat-current (ZHC) conductivity at zero density is reduced to the
electrical conductivity, the trivial flow of ZHC conductivity can be
naturally viewed as the nontrivial extension of the zero-density electrical
conductivity flow \cite{Liu0809}. Furthermore, given the analytical
expression of electric and thermoelectric conductivities that can be
obtained from the conserved electric current, we can derive the thermal
conductivity analytically by using the trivial RG flow of ZHC conductivity
and the infrared boundary condition of the matrix Riccati equation. The
radially conserved thermal current is not required.

The rest of this paper is organized as follows. In Sec. 2, we will develop a
general framework for the holographic RG flow of the thermo-electric
transport. In Sec. 3, we will take the EMA model as an example which
exhibits how the RG flow can be used to calculate the AC thermo-electric
conductivities on the boundary. The RG flow of the DC thermo-electric
conductivities will be studied in Sec. 4. By the numerical method, one can
find that the ZHC conductivity has a trivial flow in various holographic
models. This further induces an analytical expression of the DC thermal
conductivity, as will be shown in Sec. 5. In the last section, the
conclusion will be given. In two appendices, we will present the
thermodynamics on the membranes and a semi-analytical proof for the trivial
RG flow, respectively.

\section{Thermo-electric RG flow: a general framework}

One of the well-known approaches to the holographic RG is the (sliding)
membrane paradigm proposed in \cite{Liu0809}. It is technically convenient
to relate the linear response measured by the observers hovering outside the
horizon to that of the boundary theory. Such relation is also exhibited in
the Wilsonian approach to fluid/gravity duality \cite{Strominger1006}. The
flow equations obtained in \cite{Liu0809} can be retrieved as the $\beta $%
-functions of double-trace couplings by the holographic Wilsonian RG
approach which integrates out the ultraviolet geometry \cite%
{Son1009,Polchinski1010,Liu1010}. The equivalence between the membrane
paradigm and the holographic Wilsonian RG has been further discussed in \cite%
{Sin1102,Sin1109}.

Until now, the holographic RG flow of the complete thermo-electric transport
has not been studied and we will develop the previous membrane paradigm to
fill this gap. Our essential idea is to associate a positioned action with a
sliding membrane and reformulate the classical equations of motion (EOM) to
the RG flow of transport coefficients which are measured by intrinsic
observers.

In linear response, the change in the expectation value of any operator $%
O_{I}$ is assumed to be linear in the perturbing source $\phi _{I}$%
\begin{equation}
\delta \left\langle O_{I}\left( \omega \right) \right\rangle
=G_{R}^{IJ}\left( \omega \right) \phi _{I}\left( \omega \right) ,
\end{equation}%
where $G_{R}^{IJ}$ is the retarded Green's function%
\begin{equation}
G_{R}^{IJ}(\omega )\equiv -i\int_{0}^{\infty }dte^{i\omega t}\langle \left[
O_{I}(t),O_{J}(0)\right] \rangle .
\end{equation}%
In holography, by recasting the on-shell quadratic action as the form%
\begin{equation}
S_{\mathrm{os}}^{(2)}=\frac{1}{2}\int \frac{d\omega }{2\pi }\phi _{I}\left(
-\omega \right) G_{IJ}\left( \omega \right) \phi _{J}\left( \omega \right) ,
\label{onshell2}
\end{equation}%
the retarded Green's function can be extracted \cite{Son0205}, up to the
contact term \cite{Son0210}. In \cite{Tian17}, it has been shown that the
coupled perturbation equations in the bulk can be reformulated as a
matrix-form Riccati equation:%
\begin{equation}
\Gamma ^{\prime }=M-N\Gamma -\Gamma \tilde{N}-\Gamma O\Gamma .  \label{main}
\end{equation}%
Here $\Gamma _{IJ}\equiv G_{IJ}/\left( i\omega \right) $ is referred to the
canonical response function and the {matrices }$M$, $N$, $\tilde{N}$ and $O$
are {independent of perturbations}. They are the functions of radial
coordinate $r$ and the prime denotes the radial derivative. In the
following, we will translate $G_{IJ}(r)$ and hence $\Gamma _{IJ}(r)$ into
the RG flow of thermo-electric conductivities. Note that the process is
general for any theories of gravity which will be considered in this paper.

In terms of the standard AdS/CFT correspondence, the $\left( d+1\right) $%
-dimensional field theory lives on a conformal class of the asymptotic
boundary of the $\left( d+2\right) $-dimensional bulk spacetime. The radial
coordinate in the bulk can be identified with certain energy scale. As a
direct extrapolation, we assume that the field theory at certain energy
scale is associated with a fictitious membrane at the radial cutoff $r=r_{c}$%
, with the line element%
\begin{equation}
ds^{2}=\frac{1}{\Lambda (r_{c})^{2}}\gamma _{ab}(r_{c})dx^{a}dx^{b}.
\label{metric3}
\end{equation}%
Here $\gamma _{ab}$ is the induced metric, with $a,b\in \left\{ 0,\cdots
,d\right\} $. To be simple, it is assumed to be homogeneous and isotropic.
Its spatial component is denoted as $\gamma _{ij}$, with $i,j\in \left\{
1,\cdots ,d\right\} $. We define $\lambda _{ab}\equiv \gamma _{ab}/\Lambda
(r_{c})^{2}$ as the membrane metric, which is determined up to a conformal
factor $\Lambda (r_{c})^{2}$ that will be specified later.

Consider that the observers on the membranes are equipped with the proper
intrinsic coordinates,%
\begin{equation}
\hat{t}=\frac{\sqrt{-\gamma _{00}(r_{c})}}{\Lambda (r_{c})}t,\;\hat{x}^{i}=%
\frac{\sqrt{\gamma _{ii}(r_{c})}}{\Lambda (r_{c})}x^{i}.
\end{equation}%
Put differently, the intrinsic observers measure the physical quantities by
the orthonormal bases. For the sake of brevity, we will describe the
positioned physical quantities as \textquotedblleft
observed\textquotedblright\ when they are measured by the intrinsic
observers lived on the membranes. To be clear, we hat on all observed
quantities. We choose to hat the vector or tensor on the index.

We need to define the positioned on-shell action, which involves three parts%
\begin{equation}
S_{\mathrm{os}}=\left. \left( S_{\mathrm{bulk}}+S_{\mathrm{GH}}+S_{\mathrm{ct%
}}\right) \right\vert _{\mathrm{on-shell}}.
\end{equation}%
The first is the bulk action%
\begin{equation}
S_{\mathrm{bulk}}=\int_{r_{+}}^{r_{c}}d^{d+2}x\sqrt{-g}\mathcal{L}.
\end{equation}%
In the AdS/CFT correspondence, the field theory lives on the boundary and
the ultraviolet limit (that we suppose to be $r_{c}\rightarrow \infty $) is
imposed. Here we consider the bulk region from the horizon $r_{+}$ to
certain cutoff surface with $r_{c}>r_{+}$, giving rise to the $r_{c}$%
-dependence of the action. Second, to implement a well-defined variational
principle, the Gibbons-Hawking term on the cutoff surface is necessary. The
last is the counterterm, which is required in AdS/CFT to cancel the
ultraviolet divergence. To obtain a continuous RG flow, we extend the
counterterm to arbitrary slices following Ref. \cite{Sin1109}.

We proceed to define the electric current and energy-momentum current on the
membranes, which are covariant,%
\begin{equation}
J^{a}=\frac{1}{\sqrt{-\lambda }}\frac{\delta S_{\mathrm{os}}}{\delta A_{a}}%
,\;T^{ab}=\frac{2}{\sqrt{-\lambda }}\frac{\delta S_{\mathrm{os}}}{\delta
\lambda _{ab}},  \label{JT}
\end{equation}%
where $\lambda $ is the determinant of the membrane metric. For our purpose,
we set $x=x^{1}$ and focus on the relevant components $\left(
J_{x},T_{tx}\right) \,$. They are observed by%
\begin{eqnarray}
J_{\hat{x}} &=&\frac{\Lambda }{\sqrt{\gamma _{11}}}J_{x}=\frac{\sqrt{\gamma
_{11}}}{\sqrt{-\gamma _{00}}}\frac{1}{\sqrt{\lambda _{d}}}\frac{\delta S_{%
\mathrm{os}}}{\delta A_{x}},  \notag \\
T_{\hat{t}\hat{x}} &=&\frac{\Lambda }{\sqrt{\gamma _{11}}}\frac{\Lambda }{%
\sqrt{-\gamma _{00}}}T_{tx}=-\frac{\sqrt{\gamma _{11}}}{\Lambda }\frac{1}{%
\sqrt{\lambda _{d}}}\frac{\delta S_{\mathrm{os}}}{\delta \lambda _{tx}}.
\label{TtxJ}
\end{eqnarray}%
where $\delta \lambda _{xt}=\delta \lambda _{tx}$ has been used. With these
quantities at hand, the observed thermo-electric conductivities can be
defined through the generalized Ohm's law%
\begin{equation}
\left(
\begin{array}{c}
J_{\hat{x}} \\
J_{\hat{x}}^{Q}%
\end{array}%
\right) =\left(
\begin{array}{cc}
\hat{\sigma} & \hat{T}\hat{\alpha} \\
\hat{T}\hat{\alpha} & \hat{T}\hat{\kappa}%
\end{array}%
\right) \left(
\begin{array}{c}
E_{\hat{x}} \\
-\nabla _{_{\hat{x}}}\hat{T}/\hat{T}%
\end{array}%
\right) ,  \label{sakJ}
\end{equation}%
where the Tolman temperature on the membrane is determined by the Hawking
temperature and the redshift factor, that is, $\hat{T}\left( r_{c}\right) =T%
\frac{\Lambda (r_{c})}{\sqrt{-\gamma _{00}(r_{c})}}$. Note that the Tolman
temperature is the only observed thermodynamic quantity which is necessary
for calculating the observed thermo-electric conductivities. Nevertheless,
we will study in Appendix A the complete and self-consistent observed
thermodynamics, which should be important in itself.

We will relate the sources $E_{\hat{x}}$ and $\nabla _{_{\hat{x}}}\hat{T}$
to the fluctuations $\delta A_{_{\hat{x}}}$ and $\delta \lambda _{\hat{t}%
\hat{x}}$, following Sec. 2.7 in \cite{Hartnoll0903}. Consider the spacetime
associated with the metric $\lambda _{\hat{a}\hat{b}}$ that is nothing but
the Minkowski metric. Rescale the time by $\hat{t}\rightarrow \bar{t}/\hat{T}
$ and then the metric has $\lambda _{\bar{t}\bar{t}}=-1/\hat{T}^{2}$. Turn
on a small constant thermal gradient $\hat{T}\rightarrow \hat{T}-\hat{x}%
\nabla _{\hat{x}}\hat{T}$. It implies $\delta \lambda _{\bar{t}\bar{t}}=-2%
\hat{x}\nabla _{\hat{x}}\hat{T}/\hat{T}^{3}$. The fluctuation can be
compensated by the diffeomorphism $\delta \lambda _{\bar{t}\bar{t}%
}=2\partial _{\bar{t}}\xi _{\bar{t}}$ with the parameter $\xi _{\bar{t}}=i%
\hat{x}\nabla _{\hat{x}}\hat{T}/\left( \bar{\omega}\hat{T}^{3}\right) $.
Here we have endowed all quantities with a time dependence $e^{-i\bar{\omega}%
\bar{t}}$. Taking $\xi _{\hat{x}}=0$, the diffeomorphisms $\delta \lambda _{%
\bar{t}\hat{x}}=\partial _{\hat{x}}\xi _{\bar{t}}$ and $\delta A_{\hat{x}%
}=A_{\bar{t}}\partial _{\hat{x}}\xi ^{\bar{t}}$ can induce $\delta \lambda _{%
\bar{t}\hat{x}}=i\nabla _{\hat{x}}\hat{T}/\left( \bar{\omega}\hat{T}%
^{3}\right) $ and $\delta A_{\hat{x}}=-iA_{\bar{t}}\nabla _{\hat{x}}\hat{T}%
/\left( \bar{\omega}\hat{T}\right) $, respectively. Rescaling back to the
original time $\hat{t}$, one can obtain the net effect of the thermal
gradient $i\hat{\omega}\delta \lambda _{\hat{t}\hat{x}}=-\nabla _{_{\hat{x}}}%
\hat{T}/\hat{T}$ and $i\hat{\omega}\delta A_{\hat{x}}=A_{\hat{t}}\nabla _{%
\hat{x}}\hat{T}/\hat{T}$. Combined with the relation $E_{\hat{x}}=i\hat{%
\omega}\delta A_{_{\hat{x}}}$ when the electric field is turned on, we can
read%
\begin{equation}
E_{\hat{x}}+A_{\hat{t}}\nabla _{_{\hat{x}}}\hat{T}/\hat{T}=i\hat{\omega}%
\delta A_{_{\hat{x}}},\;\nabla _{_{\hat{x}}}\hat{T}/\hat{T}=-i\hat{\omega}%
\delta \lambda _{\hat{t}\hat{x}}.  \label{ET}
\end{equation}%
Furthermore, the variation of the on-shell action takes the form%
\begin{eqnarray}
\delta S_{\mathrm{os}} &=&\int d^{d+1}x\left( \frac{\delta S_{\mathrm{os}}}{%
\delta A_{x}}\delta A_{x}+\frac{\delta S_{\mathrm{os}}}{\delta \gamma _{tx}}%
\delta \gamma _{tx}\right)  \notag \\
&=&\int d^{d+1}x\sqrt{-\lambda }\left( J^{x}\delta A_{x}+T^{tx}\delta
\lambda _{tx}\right)  \notag \\
&=&\int d^{d+1}\hat{x}\sqrt{-\hat{\lambda}}\left( J^{\hat{x}}\delta A_{_{%
\hat{x}}}+T^{\hat{t}\hat{x}}\delta \lambda _{\hat{t}\hat{x}}\right)  \notag
\\
&=&\int d^{d+1}\hat{x}\sqrt{-\hat{\lambda}}\left[ J^{\hat{x}}\frac{E_{\hat{x}%
}}{i\hat{\omega}}-\left( T_{\hat{t}}^{\hat{x}}+A_{\hat{t}}J^{\hat{x}}\right)
\frac{-\nabla _{_{\hat{x}}}\hat{T}}{i\hat{\omega}\hat{T}}\right] .
\end{eqnarray}%
In the second line, we have used Eq. (\ref{JT}) and $\delta \lambda
_{tx}=\delta \gamma _{tx}/\Lambda (r_{c})^{2}$. The third line denotes a
coordinate transformation. In terms of Eq. (\ref{ET}) we obtain the last
line, where the heat current can be recognised
\begin{equation}
J_{\hat{x}}^{Q}=-\left( T_{\hat{t}\hat{x}}+A_{\hat{t}}J_{\hat{x}}\right) .
\label{JxQJ}
\end{equation}%
Putting Eq. (\ref{TtxJ}) and Eq. (\ref{JxQJ}) into Eq. (\ref{sakJ}), we can
represent $\left( \hat{\sigma},\hat{\alpha},\hat{\kappa}\right) $ by%
\begin{eqnarray}
\hat{\sigma} &=&\frac{1}{i\omega }G_{11}\frac{\gamma _{11}}{\sqrt{\lambda
_{d}}\Lambda ^{2}},  \notag \\
\hat{\alpha} &=&\frac{1}{i\omega }\frac{\left( -G_{12}\gamma
_{00}-G_{11}\gamma _{11}A_{t}\right) }{\sqrt{\lambda _{d}}\Lambda ^{2}T},
\notag \\
\hat{\kappa} &=&\frac{1}{i\omega }\frac{1}{\gamma _{11}\sqrt{\lambda _{d}}%
\Lambda \sqrt{-\gamma _{00}}T}\left[ G_{11}\gamma _{11}^{2}A_{t}^{2}+\left(
G_{12}+G_{21}\right) \gamma _{00}\gamma _{11}A_{t}+\left(
G_{22}-C_{22}\right) \gamma _{00}^{2}\right] .  \label{Ansak}
\end{eqnarray}%
Here we have defined the correlator $G_{IJ}$ by Eq. (\ref{onshell2}). The
sources $\varphi _{I}=\left( a_{x},h_{tx}\right) $ come from\footnote{%
Hereafter, we will drop the index $c$ in $r_{c}$ for brevity.}%
\begin{equation}
\delta \gamma _{tx}=\gamma _{11}(r)h_{tx}(r)e^{-i\omega t},\;\delta
A_{x}=a_{x}(r)e^{-i\omega t}.
\end{equation}%
It should be noted that the contact term $C_{22}\equiv G_{22}\left( 0\right)
$ (that appears in all the models of this paper) has been subtracted in Eq. (%
\ref{Ansak}), otherwise there is a pole at $\omega =0$ in the imaginary part
of $\hat{\kappa}$ \cite{Sin1409}. The observed ZHC conductivity is defined by%
\begin{equation}
\hat{\sigma}_{\mathrm{0}}\equiv \left. \frac{J_{\hat{x}}}{E_{\hat{x}}}%
\right\vert _{J_{\hat{x}}^{Q}=0}=\hat{\sigma}-\frac{\hat{T}\hat{\alpha}^{2}}{%
\hat{\kappa}}.
\end{equation}%
From Eq. (\ref{Ansak}), it can be expressed as%
\begin{equation}
\hat{\sigma}_{\mathrm{0}}=\frac{1}{i\omega }\frac{-\gamma _{00}G_{11}}{\sqrt{%
\lambda _{d}}\Lambda ^{2}}\frac{G_{11}\left( G_{12}-G_{21}\right) \gamma
_{11}A_{t}-\left[ G_{11}\left( G_{22}-C_{22}\right) -G_{12}^{2}\right]
\gamma _{00}}{G_{11}A_{t}^{2}\gamma _{11}^{2}+\left( G_{12}+G_{21}\right)
\gamma _{00}\gamma _{11}A_{t}+\left( G_{22}-C_{22}\right) \gamma _{00}^{2}}.
\label{OC}
\end{equation}

We need to specify the conformal factor $\Lambda ^{2}$. In order for the RG
flow to meet the AdS/CFT on the boundary, the conformal factor should have $%
\Lambda ^{2}\rightarrow $ $\gamma _{11}$ as $r\rightarrow \infty $. To
determine it completely, we note that the definition of the membrane
electric conductivity in the first line of Eq. (\ref{Ansak}) is different
from that in \cite{Liu0809}, which is $G_{11}/\left( i\omega \right) $. The
difference comes from three aspects: i) our currents (\ref{JT}) are
covariant on the membrane; ii) our physical quantities are measured by the
intrinsic observer; iii) we have rescaled the induced metric. Except the
last one, our formulation is close to Ref. \cite%
{Wilczek97,Strominger1006,Tian1407}, which treat the membrane as an
effective physical system, so the physical quantities should be more
suitably defined as intrinsic tensor (vector, scalar) fields and measured by
the intrinsic observer. However, the difference might not be substantial,
since it can be removed by a simple scaling transformation on the membrane
(at least when it is homogeneous and isotropic). Moreover, the definition in
\cite{Liu0809} is interesting at least because its flow (with zero charge
density) does not run. Keeping these in mind, we can require both
definitions to be consistent by conveniently selecting the conformal factor
as%
\begin{equation}
\Lambda ^{2}=\frac{\gamma _{11}}{\sqrt{\lambda _{d}}}=\gamma _{11},
\end{equation}%
where the isotropy has been imposed.

\section{AC thermo-electric conductivities on the boundary}

A simple holographic framework with momentum relaxation was presented in
\cite{Andrade1311}. The model contains linear axions $\chi _{i}$ along
spatial directions. We consider the four-dimensional EMA theory described by
the bulk action%
\begin{equation}
S_{\mathrm{bulk}}=\int d^{4}x\sqrt{-g}\left[ R+6-\frac{1}{4}F^{2}-\frac{1}{2}%
\sum\limits_{i=1}^{2}\left( \partial \chi _{i}\right) ^{2}\right] .
\label{bulkEMA}
\end{equation}%
Here the AdS radius $L$ and the Newton constant $16\pi G_{N}$ are set to
unity. The EMA theory allows a (homogeneous and isotropic) black-brane
solution:%
\begin{eqnarray}
ds^{2} &=&-h(r)dt^{2}+\frac{1}{h(r)}dr^{2}+r^{2}(dx_{1}^{2}+dx_{2}^{2}),
\notag \\
h(r) &=&r^{2}-\frac{r_{+}^{3}}{r}-(1-\frac{r_{+}}{r})\left( \frac{r_{+}}{4r}%
\mu ^{2}+\frac{1}{2}\beta ^{2}\right) ,  \notag \\
A &=&\mu (1-\frac{r_{+}}{r})dt,\;\chi _{i}=\beta x_{i}.  \label{EMA metric}
\end{eqnarray}%
The Hawking temperature and the charge density can be read off:%
\begin{equation}
T=\frac{1}{4\pi }\left( 3r_{+}-\frac{\beta ^{2}}{2r_{+}}-\frac{q^{2}}{%
4r_{+}^{3}}\right) ,\;q=\mu r_{+}.  \label{Ts}
\end{equation}

Perturb the background by the vector modes along $x=x_{1}$ direction, which
we write as%
\begin{equation}
\delta g_{tx}=r^{2}h_{tx}(r)e^{-i\omega t},\;\delta
A_{x}=a_{x}(r)e^{-i\omega t},\;\delta \chi _{1}=\beta ^{-1}\chi \left(
r\right) e^{-i\omega t}.
\end{equation}%
The relevant EOM are%
\begin{eqnarray}
\left( qh_{tx}+ha_{x}^{\prime }\right) ^{\prime }+\frac{\omega ^{2}}{h}a_{x}
&=&0,  \notag \\
\left( r^{2}h\chi ^{\prime }\right) ^{\prime }-\frac{i\omega r^{2}}{h}\left(
\beta ^{2}h_{tx}+i\omega \chi \right) &=&0,  \notag \\
\chi ^{\prime }-\frac{i\omega }{r^{2}h}\left( qa_{x}+r^{4}h_{tx}^{\prime
}\right) &=&0.  \label{EOMEMA}
\end{eqnarray}%
By setting $\psi \equiv r^{2}h\chi ^{\prime }/\omega $ one can reduce the
EOM to%
\begin{eqnarray}
\left( ha_{x}^{\prime }\right) ^{\prime } &=&A_{11}a_{x}+A_{12}\psi ,  \notag
\\
\left( r^{-2}h\psi ^{\prime }\right) ^{\prime } &=&A_{21}a_{x}+A_{22}\psi ,
\label{EOM2}
\end{eqnarray}%
where%
\begin{equation}
A=\left(
\begin{array}{cc}
\frac{\omega ^{2}}{h}-\frac{q^{2}}{r^{4}} & -\frac{iq}{r^{4}} \\
\beta ^{2}\frac{iq}{r^{4}} & \frac{1}{r^{2}}(\frac{\omega ^{2}}{h}-\frac{%
\beta ^{2}}{r^{2}})%
\end{array}%
\right) .
\end{equation}%
Now we will reformulate the EOM (\ref{EOM2}) as a matrix-form Riccati
equation. Define an auxiliary transport matrix $\tau $ by%
\begin{equation}
\left(
\begin{array}{c}
-ha_{x}^{\prime } \\
-r^{-2}h\psi ^{\prime }%
\end{array}%
\right) =\left(
\begin{array}{cc}
\tau _{11} & \tau _{12} \\
\tau _{21} & \tau _{22}%
\end{array}%
\right) \left(
\begin{array}{c}
i\omega a_{x} \\
i\omega \psi%
\end{array}%
\right) .  \label{matrix1}
\end{equation}%
It is different from the canonical response function $\Gamma $. We adapt
this non-canonical representation since the numerical calculation is more
simple. We stress that $\tau $ is required to be regular on the horizon by
the suitable selection of the left hand side in Eq. (\ref{matrix1}). After a
little matrix calculation, one can obtain the radial evolution equation%
\begin{equation}
\tau ^{\prime }=\frac{1}{i\omega }A+i\omega \tau B\tau ,  \label{matrixEOM}
\end{equation}%
where%
\begin{equation}
B=\frac{1}{h}\left(
\begin{array}{cc}
1 & 0 \\
0 & r^{2}%
\end{array}%
\right) .
\end{equation}%
The simple equation (\ref{matrixEOM}) is a matrix-form Riccati equation
which has been derived previously in \cite{Tian17}. It should be noted that
a key technique to build up Eq. (\ref{matrixEOM}) is to introduce two
auxiliary modes $\tilde{a}_{x}$ and $\tilde{\psi}$ to double two $2\times 1$
matrix in Eq. (\ref{matrix1}) as two $2\times 2$ matrix. Then the matrix
manipulation is fluent.

Applying the regularity of $\tau $ on the horizon, we read off the horizon
value of $\tau $ from (\ref{matrixEOM}):
\begin{equation}
\tau (r_{+})=\left(
\begin{array}{cc}
1 & 0 \\
0 & \frac{1}{r_{+}^{2}}%
\end{array}%
\right) .  \label{bc}
\end{equation}%
Taking $\tau (r_{+})$ as the boundary condition, the flow $\tau (r)$ can be
integrated out.

We write down the Gibbons-Hawking term and the counterterm \cite{Sin1409}%
\begin{eqnarray}
S_{\mathrm{GH}} &=&-2\int d^{3}x\sqrt{-\gamma }K,  \label{GHEMA} \\
S_{\mathrm{ct}} &=&\int d^{3}x\sqrt{-\gamma }\left( -4+\frac{1}{2}%
\sum\limits_{i=1}^{2}\gamma ^{ab}\partial _{a}\chi _{i}\partial _{b}\chi
_{i}\right) ,  \label{ctEMA}
\end{eqnarray}%
where $K$ is the external curvature. Then we have the positioned on-shell
action $S_{\mathrm{os}}$, from which we can calculate the one-point functions%
\footnote{%
In this paper, we neglect the terms $\sim \chi $ in all one-point functions.
They do not affect the thermo-electric conductivities.}%
\begin{eqnarray}
\frac{\delta S_{\mathrm{os}}}{\delta a_{x}} &=&-qh_{tx}-ha_{x}^{\prime },
\notag \\
\frac{\delta S_{\mathrm{os}}}{\delta h_{tx}} &=&r^{4}h_{tx}^{\prime }+\bar{C}%
_{22}h_{tx}.  \label{onepoint}
\end{eqnarray}%
Here we have defined a real radial function%
\begin{equation}
\bar{C}_{22}=4r^{3}\left( 1-\frac{r}{\sqrt{h}}\right) .  \label{C22bEMA}
\end{equation}%
Its details is useful only in Appendix B. Applying Eq. (\ref{EOMEMA}) and
Eq. (\ref{matrix1}) to eliminate the derivatives of sources in Eq. (\ref%
{onepoint}), we can obtain%
\begin{eqnarray}
G_{11} &=&i\omega \left( \tau _{11}-\frac{\tau _{12}\tau _{21}}{\tau _{22}}%
\right) ,\;G_{12}=-\left( i\beta ^{2}\frac{\tau _{12}}{\tau _{22}}+q\right) ,
\notag \\
G_{21} &=&\left( \frac{i\tau _{21}}{\tau _{22}}-q\right) ,\;G_{22}=\bar{C}%
_{22}+\frac{i\beta ^{2}}{\omega \tau _{22}}.  \label{GTau}
\end{eqnarray}%
One can see that $\bar{C}_{22}$ is part of the contact term $C_{22}\equiv
G_{22}\left( 0\right) $. Inserting Eq. (\ref{GTau}) into Eq. (\ref{Ansak})
with $\gamma _{00}=-h$, $\gamma _{11}=r^{2}$, and $\lambda _{d}=1$, $\left(
\hat{\sigma},\hat{\alpha},\hat{\kappa}\right) $ can be related to $\tau $.
For instance,%
\begin{equation}
\hat{\sigma}=\frac{1}{i\omega }G_{11}=\tau _{11}-\frac{\tau _{12}\tau _{21}}{%
\tau _{22}}.  \label{sjG11}
\end{equation}

\begin{figure}[th]
\centerline{
\includegraphics[width=1\textwidth]{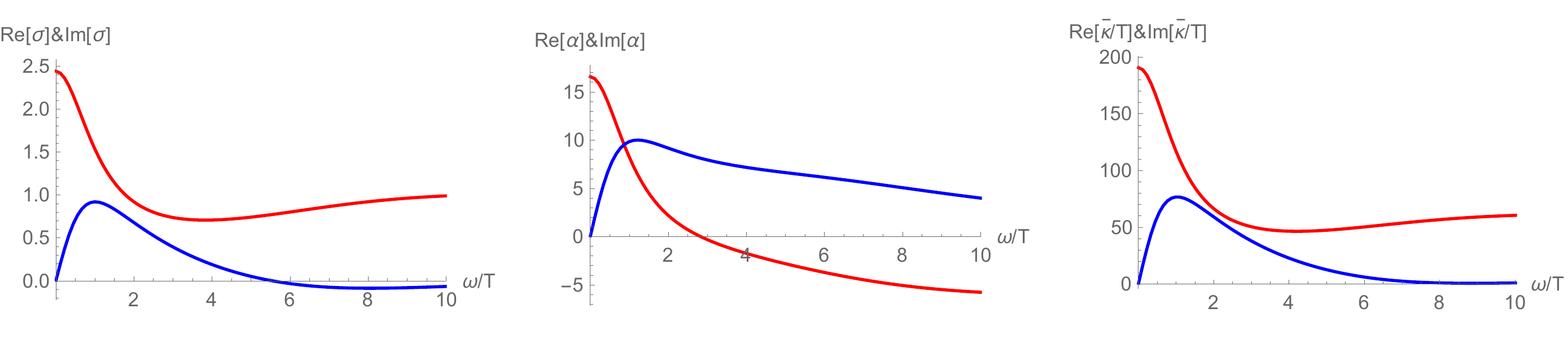}}
\caption{AC thermo-electric conductivities in the EMA model. The red and
blue curves denote real and imaginary parts, respectively. We plot $\left(
\hat{\protect\sigma},\hat{\protect\alpha},\hat{\protect\kappa}\right) $ as
the functions of $\protect\omega /T$ on the boundary $r_{+}/r=10^{-4}$. We
fix the dimensionless parameters $\protect\mu /T=6$ and $\protect\beta /T=5$
in order to compare with the green curves given in Figure 2 and Figure 8 in
Ref. \protect\cite{Sin1409}.}
\label{AC}
\end{figure}

Now we can implement the numerical calculation and plot the AC
thermo-electric conductivities. We focus on the limit $r\rightarrow \infty $%
, see Figure \ref{AC}. They are denoted by $\left( \sigma ,\alpha ,\bar{%
\kappa}\right) $. The results are same to Ref. \cite{Sin1409}. Note that we
fix $r_{+}=1$ in all numerical calculations of this paper.

\section{RG flow of ZHC conductivity in the DC limit}

It is direct to show numerically that the RG flow of ZHC conductivity does
not run in the DC limit. This is what we will do in the following for
various holographic models. In Appendix B, we will present an alternative
semi-analytical method. As a bonus, we will obtain the analytical expression
of the contact term.

\subsection{Einstein-Maxwell-axion model}

In Figure 2, we plot the trivial line that describes $\hat{\sigma}_{0}(r)$
and compare it with the nontrivial RG flow of three thermo-electric
conductivities. It is amazing that the nontrivial evolution of $\left( \hat{%
\sigma},\hat{\alpha},\hat{\kappa}\right) $ exactly cancels each other to
produce $\hat{\sigma}_{0}(r)=\mathrm{const}$.
\begin{figure}[th]
\centerline{
\includegraphics[width=1\textwidth]{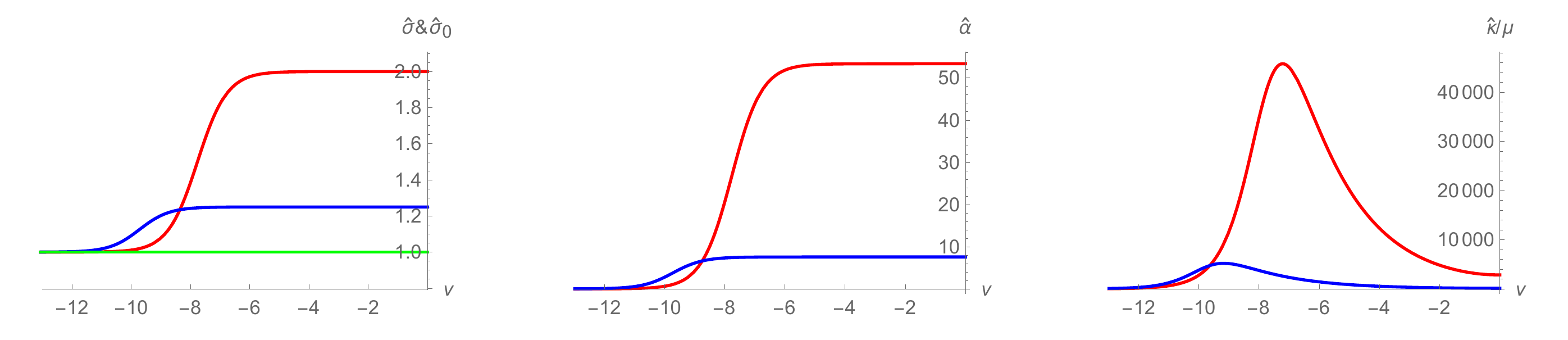}}
\caption{$\left( \hat{\protect\sigma},\hat{\protect\alpha},\hat{\protect%
\kappa}\right) $ and $\hat{\protect\sigma}_{\mathrm{0}}$ as the functions of
$v=\log (1-r_{+}/r)$ at $\protect\omega /\protect\mu =10^{-4}$ in the EMA
model. The coordinate $v$ is used to highlight the near-horizon behavior.
For $\left( \hat{\protect\sigma},\hat{\protect\alpha},\hat{\protect\kappa}%
\right) $, we fix the parameters $T/\protect\mu =1$ and $\protect\beta /%
\protect\mu =1$ (red), or $T/\protect\mu =1/2$ and $\protect\beta /\protect%
\mu =2$ (blue). The green lines depict $\hat{\protect\sigma}_{\mathrm{0}}$
for both groups of parameters. They merge in the EMA model. But it is not
the case for other models.}
\label{OCEMA}
\end{figure}

\subsection{Gauss-Bonnet curvature}

The higher derivative corrections appear generally in any quantum gravity
theory from quantum or stringy effects. These corrections may be holographic
dual to $1/N$ or $1/\lambda $ corrections in some gauge theories, allowing
independent values of two central charges $a$ and $c$. This is in contrast
to the standard $\mathcal{N}$=4 super Yang-Mills theory where $a=c$.
Actually, the Gauss-Bonnet (GB) correction has been treated as a dangerous
source of violation for the feature that is universal in the Einstein
gravity \cite{Liu1305}. In the following, we will use GB gravity as a good
test for the universality of RG flow.

Consider the GB correction to the EMA theory, with the bulk action \cite%
{Brihaye2008,Astefanesei0806,Liu0807,Ge1411GB}
\begin{eqnarray}
S_{\mathrm{bulk}} &=&\int d^{5}x\sqrt{-g}\big[R+12-\frac{1}{4}F^{\mu \nu
}F_{\mu \nu }-\frac{1}{2}\sum\limits_{i=1}^{3}\left( \partial \chi
_{i}\right) ^{2}  \notag \\
&&+\frac{\tilde{\alpha}}{2}\left( R^{2}-4R^{\mu \nu }R_{\mu \nu }+R_{\mu \nu
\lambda \rho }R^{\mu \nu \lambda \rho }\right) \big],  \label{bulkGB}
\end{eqnarray}%
where $\tilde{\alpha}$ is the GB coupling constant\footnote{%
Without the axions, there exists a constraint $-\frac{7}{36}\leq \tilde{%
\alpha}\leq \frac{9}{100}$ by requiring the causality of field theories on
the boundary \cite{Brigante08} or the positivity of the energy flux \cite%
{Hofman08}. Moreover, it has been pointed out that any nonzero $\tilde{\alpha%
}$ requires an infinite number of massive higher spin fields to respect the
causality \cite{Maldacena1407}. But see \cite{Papallo1508} for different
arguments. The disorder parameter can also affect the causality \cite{Ge1605}%
.}. The isotropic black-brane solution can be written as \cite{Ge1411GB}%
\begin{eqnarray}
ds^{2} &=&-h(r)dt^{2}+\frac{1}{f(r)}%
dr^{2}+r^{2}(dx_{1}^{2}+dx_{2}^{2}+dx_{3}^{2}),  \notag \\
f(r) &=&\frac{r^{2}}{2\tilde{\alpha}}\left[ 1-\sqrt{1+\tilde{\alpha}\frac{%
2\left( r^{2}-r_{+}^{2}\right) }{3r^{6}r_{+}^{2}}\left[
q^{2}+3r^{2}r_{+}^{2}\left( \beta ^{2}-4r^{2}-4r_{+}^{2}\right) \right] }%
\right] ,  \notag \\
h(r) &=&L_{\mathrm{eff}}^{2}f(r),\;A=\frac{L_{\mathrm{eff}}q}{2r_{+}^{2}}(1-%
\frac{r_{+}^{2}}{r^{2}})dt,\;\chi _{i}=\beta x_{i},\;i=1,2,3.
\label{GB metric}
\end{eqnarray}%
Here $L_{\mathrm{eff}}^{2}=\frac{1+\sqrt{1-4\tilde{\alpha}}}{2}$ is the
square of the effective AdS radius. In contrast to the GB metric that is
usually used in the literature, we have rescaled $t\rightarrow tL_{\mathrm{%
eff}}$. Thus, the RG flow that we will construct can match on the boundary
to the AdS/CFT result. For instance, the observed temperature is $\hat{T}%
\left( r\right) =\frac{r}{\sqrt{h(r)}}T$, which can be directly reduced to
the Hawking temperature $T$ at $r\rightarrow \infty $ due to our rescaling
of time. The temperature and charge density can be written as%
\begin{equation}
T=\frac{L_{\mathrm{eff}}}{\pi }\left( r_{+}-\frac{\beta ^{2}}{8r_{+}}-\frac{%
q^{2}}{24r_{+}^{5}}\right) ,\;q=2r_{+}^{2}\frac{\mu }{L_{\mathrm{eff}}}.
\label{GBTsro}
\end{equation}

Consider the relevant modes along $x=x_{1}$ direction, which are given by%
\begin{equation}
\delta g_{tx}=r^{2}h_{tx}(r)e^{-i\omega t},\;\delta
A_{x}=a_{x}(r)e^{-i\omega t},\;\delta \chi _{1}=\beta ^{-1}\chi \left(
r\right) e^{-i\omega t}.
\end{equation}%
Define an auxiliary transport matrix $\tau $ by%
\begin{equation}
\left(
\begin{array}{c}
-r\sqrt{hf}a_{x}^{\prime } \\
-r^{-3}\sqrt{hf}\psi ^{\prime }%
\end{array}%
\right) =\left(
\begin{array}{cc}
\tau _{11} & \tau _{12} \\
\tau _{21} & \tau _{22}%
\end{array}%
\right) \left(
\begin{array}{c}
i\omega a_{x} \\
i\omega \psi%
\end{array}%
\right) ,  \label{MatrixGB}
\end{equation}%
where $\psi =r^{3}\sqrt{hf}\chi ^{\prime }/\omega $. From three EOM%
\begin{eqnarray}
\left( qh_{tx}+r\sqrt{fh}a_{x}^{\prime }\right) ^{\prime }+\frac{r}{\sqrt{fh}%
}\omega ^{2}a_{x} &=&0,  \notag \\
\chi ^{\prime }-\frac{i\omega }{h}\left[ \left( r^{2}-2\tilde{\alpha}%
f\right) h_{tx}^{\prime }+A_{t}^{\prime }a_{x}\right] &=&0,  \notag \\
\left( r^{3}\sqrt{fh}\chi ^{\prime }\right) ^{\prime }+\frac{\omega r^{3}}{%
\sqrt{fh}}\left( \omega \chi -i\beta ^{2}h_{tx}\right) &=&0,  \label{EOMGB}
\end{eqnarray}%
one can construct a matrix-form Riccati equation%
\begin{equation}
\tau ^{\prime }=\frac{1}{i\omega }A+i\omega \tau B\tau ,  \label{GBmatrixEOM}
\end{equation}%
where the matrix $A$ and $B$ are%
\begin{equation}
A=\frac{1}{r^{3}\sqrt{fh}}\left(
\begin{array}{cc}
r^{4}\omega ^{2}-\frac{q^{2}h}{r^{2}-2\tilde{\alpha}f} & -\frac{iqh}{r^{2}-2%
\tilde{\alpha}f} \\
\frac{iq\beta ^{2}h}{r^{2}-2\tilde{\alpha}f} & \omega ^{2}-\frac{\beta ^{2}h%
}{r^{2}-2\tilde{\alpha}f}%
\end{array}%
\right) ,\;B=\frac{1}{r\sqrt{fh}}\left(
\begin{array}{cc}
1 & 0 \\
0 & r^{4}%
\end{array}%
\right) .
\end{equation}%
Applying the regularity of $\tau $ on the horizon, one can extract the
horizon value of $\tau $ from Eq. (\ref{GBmatrixEOM}) directly:
\begin{equation}
\tau (r_{+})=\left(
\begin{array}{cc}
r_{+} & 0 \\
0 & r_{+}^{-3}%
\end{array}%
\right) .
\end{equation}%
Using $\tau (r_{+})$ as the boundary condition, we can integrate out the RG
flow $\tau (r)$.

To obtain the positioned on-shell action $S_{\mathrm{os}}$, we need the
Gibbons-Hawking term and the counterterm \cite%
{Brihaye2008,Astefanesei0806,Liu0807,Ge1411GB}%
\begin{eqnarray}
S_{\mathrm{GH}} &=&-2\int d^{4}x\sqrt{-\gamma }K,  \label{GBGH} \\
S_{\mathrm{ct}} &=&\int d^{4}x\sqrt{-\gamma }\left( -6+\frac{1}{2}%
\sum\limits_{i=1}^{3}\gamma ^{ab}\partial _{a}\chi _{i}\partial _{b}\chi
_{i}\right) .  \label{ctGB}
\end{eqnarray}%
Taking the variation of $S_{\mathrm{os}}$, one can calculate the one-point
functions%
\begin{eqnarray}
\frac{\delta S_{\mathrm{os}}}{\delta a_{x}} &=&-qh_{tx}-r\sqrt{fh}%
a_{x}^{\prime },  \notag \\
\frac{\delta S_{\mathrm{os}}}{\delta h_{tx}} &=&r^{3}\sqrt{\frac{f}{h}}%
\left( r^{2}-2\tilde{\alpha}f\right) h_{tx}^{\prime }+\bar{C}_{22}h_{tx},
\label{onepointGB}
\end{eqnarray}%
where we have neglected some terms that do not contribute to the DC
conductivities. The function $\bar{C}_{22}$ is given by%
\begin{equation}
\bar{C}_{22}=\frac{r^{2}}{4L_{\mathrm{eff}}\sqrt{h}}\left[ \beta ^{2}L_{%
\mathrm{eff}}^{2}r-8r^{3}\left( 2+\sqrt{1-4\tilde{\alpha}}\right) +8L_{%
\mathrm{eff}}\sqrt{f}\left( 3r^{2}-2\tilde{\alpha}f\right) \right] .
\label{C22bGB}
\end{equation}

\begin{figure}[th]
\centerline{
\includegraphics[width=1\textwidth]{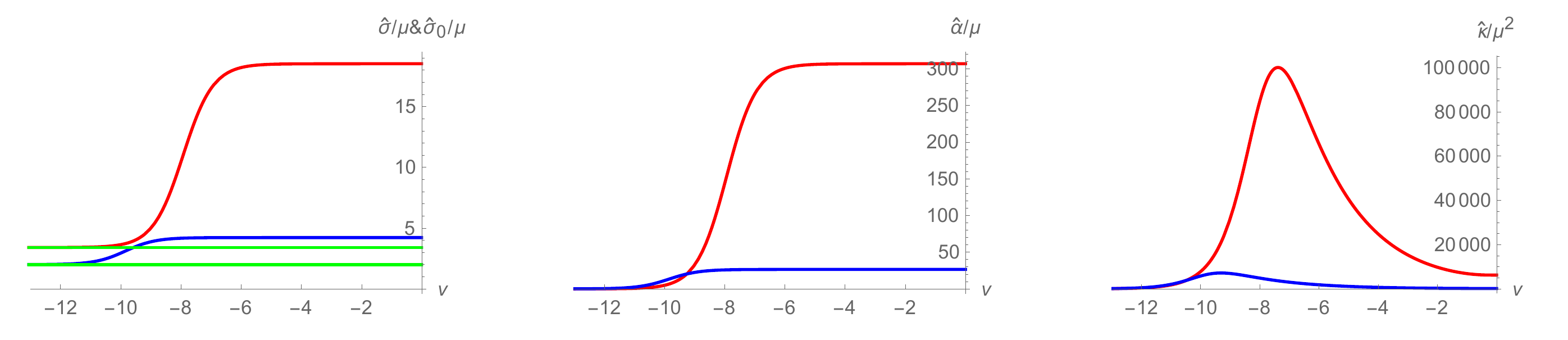}}
\caption{The RG flow of $\left( \hat{\protect\sigma},\hat{\protect\alpha},%
\hat{\protect\kappa}\right) $ and $\hat{\protect\sigma}_{\mathrm{0}}$ in the
GB model. The coupling constant is $\tilde{\protect\alpha}=9/100$. In this
and other remaining figures, the parameters $T$ and $\protect\beta $ and the
color scheme are same as those in Figure \protect\ref{OCEMA}.}
\label{OCGB}
\end{figure}

Applying Eq. (\ref{MatrixGB}) and Eq. (\ref{EOMGB}) to eliminate the
derivatives of sources in Eq. (\ref{onepointGB}), we can obtain%
\begin{eqnarray}
G_{11} &=&i\omega \left( \tau _{11}-\frac{\tau _{12}\tau _{21}}{\tau _{22}}%
\right) ,\;G_{12}=-\left( i\beta ^{2}\frac{\tau _{12}}{\tau _{22}}+q\right) ,
\notag \\
G_{21} &=&\frac{i\tau _{21}}{\tau _{22}}-q,\;G_{22}=\bar{C}_{22}+\frac{%
i\beta ^{2}}{\omega \tau _{22}}.
\end{eqnarray}%
Interestingly, they have the same form as Eq. (\ref{GTau}), up to $\bar{C}%
_{22}$. Inserting them into Eq. (\ref{Ansak}) and Eq. (\ref{OC}), $\left(
\hat{\sigma},\hat{\alpha},\hat{\kappa}\right) $ and $\hat{\sigma}_{\mathrm{0}%
}$ can be inferred from $\tau $. We plot their RG flow in Figure \ref{OCGB}.

\subsection{Dilaton field}

Adding the dilaton is natural from the dimensional reductions of consistent
string theory. In the AdS/CMT duality, the dilaton theory is particularly
appealing as it provides various distinctive physical properties \cite%
{Charmousis1005}. We will consider an Einstein-Maxwell-Axion-Dilaton (EMAD)
theory. Its bulk action is given by%
\begin{equation}
S_{\mathrm{bulk}}=\int d^{4}x\sqrt{-g}{\bigg [}R-\frac{Z\left( \phi \right)
}{4}F^{\mu \nu }F_{\mu \nu }-\frac{1}{2}\nabla _{\mu }\chi _{i}\nabla ^{\mu
}\chi _{i}-\frac{1}{2}\nabla _{\mu }\phi \nabla ^{\mu }\phi +V\left( \phi
\right) {\bigg ]},
\end{equation}%
where the gauge field coupling and the scalar potential are taken as \cite%
{Gubser0911},%
\begin{equation}
Z\left( \phi \right) =\exp (\phi /\sqrt{3}),\;V\left( \phi \right) =6\cosh
(\phi /\sqrt{3}).  \label{ZV}
\end{equation}%
In Ref. \cite{Zaanen1311}, the Einstein-Maxwell-Dilaton theory with the
massive graviton has been studied. Using Eq. (\ref{ZV}), the analytical
black brane solution has been found. We notice that there is a similar black
brane solution in the EMAD theory%
\begin{eqnarray}
ds^{2} &=&r^{2}f(r)\left( -h(r)dt^{2}+dx^{2}+dy^{2}\right) +\frac{1}{%
r^{2}f(r)h(r)}dr^{2},  \label{EMADmetric} \\
h(r) &=&1-\frac{1}{\left( Q+r\right) ^{3}}\left( m+Q^{3}+\frac{\beta ^{2}r}{2%
}\right) ,\;f(r)=\left( 1+\frac{Q}{r}\right) ^{\frac{3}{2}}  \notag \\
A &=&\frac{\sqrt{3Q\left( m+Q^{3}-\frac{\beta ^{2}}{2}\right) }}{Q+r_{+}}%
\frac{r-r_{+}}{Q+r}dt,\;\phi (r)=\frac{\sqrt{3}}{2}\log \left( 1+\frac{Q}{r}%
\right) ,\;\chi _{i}=\beta x_{i},  \notag
\end{eqnarray}%
where $m$ and $Q$ are two parameters. The temperature, chemical potential,
and charge density can be written as%
\begin{equation}
T=\frac{\sqrt{r_{+}}\left[ 6\left( Q+r_{+}\right) ^{2}-\beta ^{2}\right] }{%
8\pi \left( Q+r_{+}\right) ^{3/2}},\;\mu =\frac{q}{Q+r_{+}},\;q=\frac{\sqrt{%
3Q\left( m+Q^{3}-\frac{\beta ^{2}}{2}\right) }}{Q+r_{+}}.
\end{equation}

Next, we will derive the EOM of vector modes and build up the Riccati
equation. Note that a general EMAD model would allow various background
solutions. For the potential application in the future, we will set a
general metric ansatz%
\begin{equation}
ds^{2}=g_{tt}\left( r\right) dt^{2}+g_{rr}\left( r\right)
dt^{2}+g_{xx}\left( r\right) \left( dx^{2}+dy^{2}\right) .  \label{gmetric}
\end{equation}%
Consider the perturbation modes
\begin{equation}
\delta g_{tx}=g_{xx}\left( r\right) h_{tx}(r)e^{-i\omega t},\;\delta
A_{x}=a_{x}(r)e^{-i\omega t},\;\delta \chi _{1}=\beta ^{-1}\chi \left(
r\right) e^{-i\omega t}.
\end{equation}%
The relevant EOM are%
\begin{eqnarray}
\left( qh_{tx}+\sqrt{-\frac{g_{tt}}{g_{rr}}}Za_{x}^{\prime }\right) ^{\prime
}+\sqrt{-\frac{g_{rr}}{g_{tt}}}\omega ^{2}Za_{x} &=&0,  \notag \\
\left( \sqrt{-\frac{g_{tt}}{g_{rr}}}g_{xx}\chi ^{\prime }\right) ^{\prime
}-i\omega \sqrt{-\frac{g_{rr}}{g_{tt}}}g_{xx}\left( \beta ^{2}h_{tx}+i\omega
\chi \right) &=&0,  \notag \\
\chi ^{\prime }+i\omega \left( \frac{g_{xx}}{g_{tt}}h_{tx}^{\prime }-\sqrt{-%
\frac{g_{rr}}{g_{tt}}}\frac{1}{g_{xx}}qa_{x}\right) &=&0.  \label{EOMEMAD}
\end{eqnarray}%
They imply a matrix-form Riccati equation%
\begin{equation}
\tau ^{\prime }=\frac{1}{i\omega }A+i\omega \tau B\tau ,
\end{equation}%
where the $\tau $ matrix is defined by%
\begin{equation}
\left(
\begin{array}{c}
-\sqrt{-\frac{g_{tt}}{g_{rr}}}Za_{x}^{\prime } \\
-\sqrt{-\frac{g_{tt}}{g_{rr}}}\frac{1}{g_{xx}}\psi ^{\prime }%
\end{array}%
\right) =\left(
\begin{array}{cc}
\tau _{11} & \tau _{12} \\
\tau _{21} & \tau _{22}%
\end{array}%
\right) \left(
\begin{array}{c}
i\omega a_{x} \\
i\omega \psi%
\end{array}%
\right) ,  \label{MatrixEMAD}
\end{equation}%
and%
\begin{equation}
A=\frac{\sqrt{-g_{tt}g_{rr}}}{g_{xx}^{2}}\left(
\begin{array}{cc}
-q^{2}-\frac{g_{xx}^{2}}{g_{tt}}Z\omega ^{2} & -iq \\
i\beta ^{2}q & -\beta ^{2}-\frac{g_{xx}}{g_{tt}}\omega ^{2}%
\end{array}%
\right) ,\;B=\sqrt{-\frac{g_{rr}}{g_{tt}}}\left(
\begin{array}{cc}
\frac{1}{Z} & 0 \\
0 & g_{xx}%
\end{array}%
\right) .
\end{equation}%
On the horizon, the regularity of $\tau $ induces%
\begin{equation}
\tau (r_{+})=\left. \left(
\begin{array}{cc}
Z & 0 \\
0 & \frac{1}{g_{xx}}%
\end{array}%
\right) \right\vert _{r=r_{+}}.
\end{equation}

\begin{figure}[th]
\centerline{
\includegraphics[width=1\textwidth]{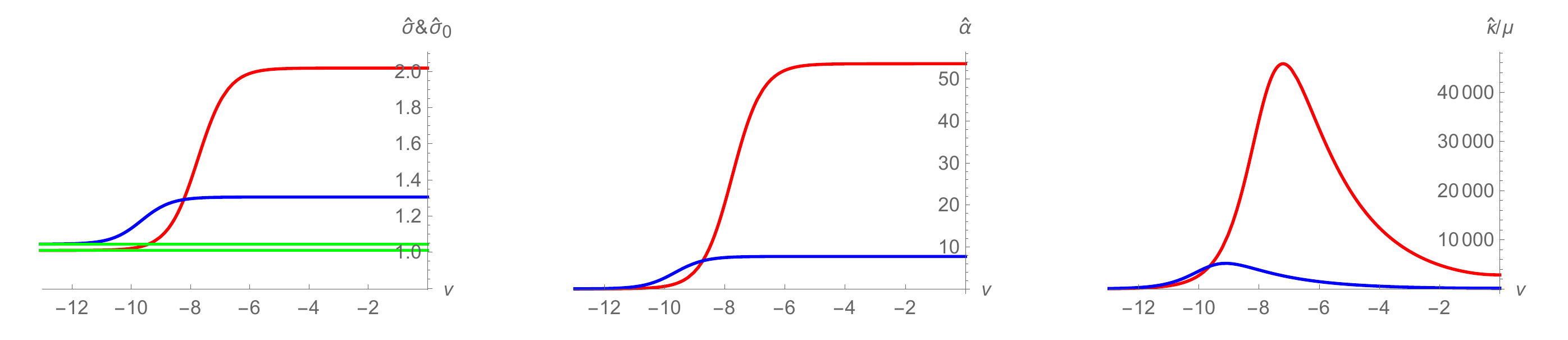}}
\caption{The RG flow of $\left( \hat{\protect\sigma},\hat{\protect\alpha},%
\hat{\protect\kappa}\right) $ and $\hat{\protect\sigma}_{\mathrm{0}}$ in the
EMAD model.}
\label{OCEMAD}
\end{figure}

The holographic renormalization of the Einstein-Maxwell-Dilaton model given
in \cite{Gubser0911} has been studied recently in \cite{Kim1608}.\ Adding
the axions does not lead to qualitative differences. Then we can read off
the counterterm%
\begin{equation}
S_{\mathrm{ct}}=\int d^{3}x\sqrt{-\gamma }\left( -4+\frac{1}{2}%
\sum\limits_{i=1}^{2}\gamma ^{ab}\partial _{a}\chi _{i}\partial _{b}\chi
_{i}+\frac{1}{3}\phi n^{r}\partial _{r}\phi -\frac{1}{6}\phi ^{2}\right) ,
\end{equation}%
where $n^{r}$ is the radial component of the outward unit vector normal to
the cutoff surface. Note that the Gibbons-Hawking term is same to one in the
EMA theory. As a result, we can derive the one-point functions%
\begin{eqnarray}
\frac{\delta S_{\mathrm{os}}}{\delta a_{x}} &=&-qh_{tx}-Z\sqrt{\frac{-g_{tt}%
}{g_{rr}}}a_{x}^{\prime },  \notag \\
\frac{\delta S_{\mathrm{os}}}{\delta h_{tx}} &=&g_{xx}^{2}\frac{1}{\sqrt{%
-g_{tt}g_{rr}}}h_{tx}^{\prime }+\bar{C}_{22}h_{tx},  \label{onepointEMAD}
\end{eqnarray}%
where%
\begin{equation}
\bar{C}_{22}=\frac{\partial _{r}g_{xx}^{2}}{\sqrt{-g_{tt}g_{rr}}}-\frac{%
g_{xx}^{2}}{\sqrt{-g_{tt}}}\left( 4+\frac{\phi ^{2}}{6}+\frac{\phi \phi
^{\prime }}{3\sqrt{g_{rr}}}\right) .  \label{C22bEMAD}
\end{equation}%
Applying Eq. (\ref{EOMEMAD}) and Eq. (\ref{MatrixEMAD}) to eliminate the
derivatives of sources in Eq. (\ref{onepointEMAD}), we can obtain%
\begin{eqnarray}
G_{11} &=&i\omega \left( \tau _{11}-\frac{\tau _{12}\tau _{21}}{\tau _{22}}%
\right) ,\;G_{12}=-\left( i\beta ^{2}\frac{\tau _{12}}{\tau _{22}}+q\right) ,
\notag \\
G_{21} &=&\left( \frac{i\tau _{21}}{\tau _{22}}-q\right) ,\;G_{22}=\bar{C}%
_{22}+\frac{i\beta ^{2}}{\omega \tau _{22}},
\end{eqnarray}%
which are still same to Eq. (\ref{GTau}), up to $\bar{C}_{22}$. Inserting
them into Eq. (\ref{Ansak}) and Eq. (\ref{OC}), we can deduce $\left( \hat{%
\sigma},\hat{\alpha},\hat{\kappa}\right) $ and $\hat{\sigma}_{\mathrm{0}}$
from $\tau $. Their RG flow is depicted in Figure \ref{OCEMAD}.

\subsection{Non-minimal coupling}

\begin{figure}[th]
\centerline{
\includegraphics[width=1\textwidth]{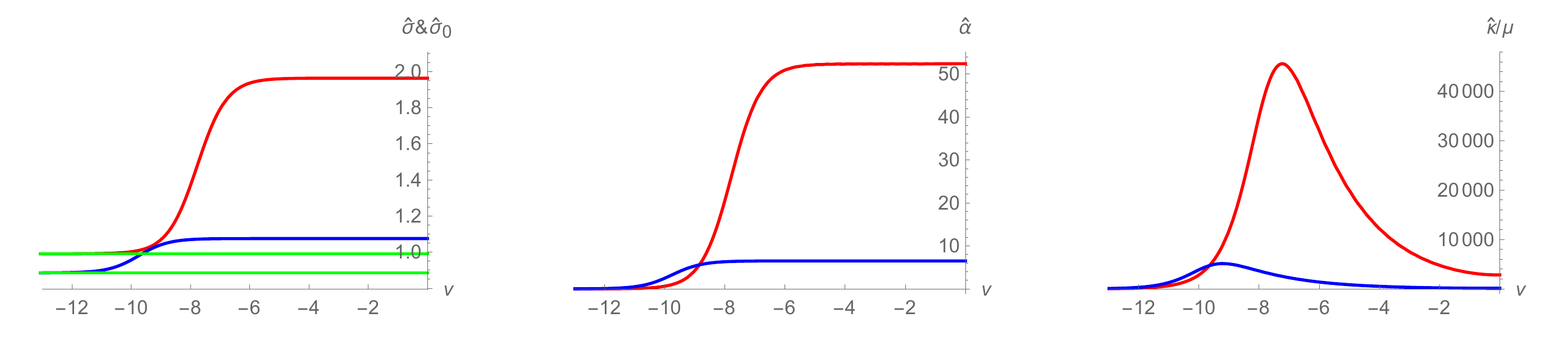}}
\caption{The RG flow of $\left( \hat{\protect\sigma},\hat{\protect\alpha},%
\hat{\protect\kappa}\right) $ and $\hat{\protect\sigma}_{\mathrm{0}}$ in the
theory with the non-minimal coupling $\mathcal{J=}2/3$.}
\label{OCNM}
\end{figure}

In all of the above models, the translation-breaking sector is minimally
coupled to the gravitational and electromagnetic sectors. There are novel
models which involve the non-minimal coupling between the Maxwell term and
the axions \cite{Baggioli1601,LWJ1602,LWJ1612}. We will focus on one of
these models, i.e. the model 1 in \cite{LWJ1602}, which has more nontrivial
conductivities than others. This model is so distinctive that it breaks
various bounds on the viscosity \cite{KSS}, electric conductivity \cite%
{Sachdev1507} and charge diffusivity \cite{Hartnoll1405}. The action is
given by
\begin{equation}
S_{\mathrm{bulk}}=\int d^{4}x\sqrt{-g}\left( R+6-\frac{1}{4}F^{2}-\frac{1}{4}%
\mathcal{J}\mathrm{Tr}\left[ \mathcal{X}F^{2}\right] -\mathrm{Tr}\left[
\mathcal{X}\right] \right) ,  \label{bulkNCAC}
\end{equation}%
where the coupling constant belongs to $0\mathcal{\leq J\leq }2/3$ by the
causality requirement and%
\begin{equation}
\mathcal{X}_{\;\nu }^{\mu }=\frac{1}{2}\sum\limits_{i=1}^{2}\partial ^{\mu
}\chi _{i}\partial _{\nu }\chi _{i}.
\end{equation}%
The background solution is same as Eq. (\ref{EMA metric}). Suppose that the
background is perturbed by the vector mode along $x=x_{1}$:%
\begin{equation}
\delta g_{tx}=r^{2}h_{tx}(r)e^{-i\omega t},\;\delta
A_{x}=a_{x}(r)e^{-i\omega t},\;\delta \chi =\beta ^{-1}\chi \left( r\right)
e^{-i\omega t}.
\end{equation}%
Due to the non-minimal coupling, the number of the relevant EOM is not three
but four:%
\begin{eqnarray}
\left( qh_{tx}+ha_{x}^{\prime }\right) ^{\prime }+\frac{\omega ^{2}}{h}a_{x}+%
\frac{2\mathcal{J}}{r(4r^{2}-\mathcal{J}\beta ^{2})}\left[ \beta ^{2}\left(
qh_{tx}+ha_{x}^{\prime }\right) +i\omega q\chi \right] &=&0,  \notag \\
\left( r^{2}h\chi ^{\prime }\right) ^{\prime }-\frac{i\omega r^{2}}{h}\left(
\beta ^{2}h_{tx}+i\omega \chi \right) -\frac{2\mathcal{J}qr}{4r^{4}+\mathcal{%
J}q^{2}}\left( i\omega \beta ^{2}a_{x}+2qh\chi ^{\prime }\right) &=&0,
\notag \\
\left( r^{4}h_{tx}^{\prime }+qa_{x}\right) ^{\prime }-\frac{r^{2}}{h}\left(
\beta ^{2}h_{tx}+i\omega \chi \right) -\frac{\mathcal{J}q}{4r^{2}h}\left[
\beta ^{2}\left( qh_{tx}+ha_{x}^{\prime }\right) +i\omega q\chi \right] &=&0,
\notag \\
\chi ^{\prime }-\frac{i\omega }{r^{2}h}\left( qa_{x}+r^{4}h_{tx}^{\prime
}\right) +\frac{i\omega \mathcal{J}q}{r^{2}h\left( 4r^{4}+\mathcal{J}%
q^{2}\right) }\left[ q\left( qa_{x}+r^{4}h_{tx}^{\prime }\right) +\beta
^{2}r^{2}a_{x}\right] &=&0.  \label{EOMNCAC}
\end{eqnarray}%
To deal with these EOM, we define a $3\times 3$ auxiliary transport matrix $%
\tau $ by%
\begin{equation}
\left(
\begin{array}{c}
-\frac{4r^{2}-\mathcal{J}\beta ^{2}}{4r^{2}}ha_{x}^{\prime } \\
r^{4}h_{tx}^{\prime } \\
-r^{2}h\chi ^{\prime }%
\end{array}%
\right) =\left(
\begin{array}{ccc}
\tau _{11} & \tau _{12} & \tau _{12} \\
\tau _{21} & \tau _{22} & \tau _{23} \\
\tau _{13} & \tau _{23} & \tau _{33}%
\end{array}%
\right) \left(
\begin{array}{c}
i\omega a_{x} \\
i\omega h_{tx} \\
i\omega \chi%
\end{array}%
\right) .  \label{tau3}
\end{equation}%
The former three EOM can be recast as%
\begin{equation}
\tau ^{\prime }=\frac{1}{i\omega }A+i\omega \tau B\tau +C\tau ,
\label{RiccatiNCAC}
\end{equation}%
where%
\begin{eqnarray}
A &=&\frac{1}{h}\left(
\begin{array}{ccc}
\frac{4r^{2}-\mathcal{J}\beta ^{2}}{4r^{2}}\omega ^{2} & \frac{\mathcal{J}%
\beta ^{2}qh}{2r^{3}} & \frac{i\omega \mathcal{J}qh}{2r^{3}} \\
0 & \frac{4r^{4}+\mathcal{J}q^{2}}{4r^{2}}\beta ^{2} & i\omega \frac{4r^{4}+%
\mathcal{J}q^{2}}{4r^{2}} \\
-\frac{2i\omega \mathcal{J}\beta ^{2}qrh}{4r^{4}+\mathcal{J}q^{2}} &
-i\omega \beta ^{2}r^{2} & r^{2}\omega ^{2}%
\end{array}%
\right) ,  \notag \\
B &=&\frac{1}{h}\left(
\begin{array}{ccc}
\frac{4r^{2}}{4r^{2}-\mathcal{J}\beta ^{2}} & 0 & 0 \\
0 & -\frac{h}{r^{4}} & 0 \\
0 & 0 & \frac{1}{r^{2}}%
\end{array}%
\right) ,\;C=\frac{q}{r^{4}}\left(
\begin{array}{ccc}
0 & \frac{4r^{2}-\mathcal{J}\beta ^{2}}{4r^{2}} & 0 \\
\frac{r^{4}}{h} & 0 & 0 \\
0 & 0 & \frac{4\mathcal{J}qr^{3}}{4r^{4}+\mathcal{J}q^{2}}%
\end{array}%
\right) .
\end{eqnarray}%
The regularity on the horizon gives%
\begin{equation}
\tau (r_{+})=\left(
\begin{array}{ccc}
\frac{4r^{2}-\mathcal{J}\beta ^{2}}{4r^{2}} & 0 & 0 \\
\frac{iq}{\omega }\frac{4r^{2}-\mathcal{J}\beta ^{2}}{4r^{2}} & \tau _{22}(r)
& \frac{i}{\omega }\frac{4r^{4}+\mathcal{J}q^{2}}{4r^{2}} \\
0 & \frac{\beta ^{2}r^{2}}{i\omega } & r^{2}%
\end{array}%
\right) _{r=r_{+}}.  \label{taurz}
\end{equation}%
To determine $\tau _{22}(r_{+})$, one can rely on the last line of Eq. (\ref%
{EOMNCAC}), which leads to the constraint%
\begin{equation}
\left(
\begin{array}{ccc}
0 & 0 & 0 \\
0 & 0 & 0 \\
0 & i\omega \frac{4r^{4}}{4r^{4}+\mathcal{J}q^{2}} & 0%
\end{array}%
\right) \left(
\begin{array}{ccc}
\tau _{11} & \tau _{12} & \tau _{12} \\
\tau _{21} & \tau _{22} & \tau _{23} \\
\tau _{13} & \tau _{32} & \tau _{33}%
\end{array}%
\right) =\left(
\begin{array}{ccc}
0 & 0 & 0 \\
0 & 0 & 0 \\
-qr^{2}\frac{4r^{2}-\mathcal{J}\beta ^{2}}{4r^{4}+\mathcal{J}q^{2}} & 0 & 0%
\end{array}%
\right) .
\end{equation}%
Combining the above two equations, one can obtain%
\begin{equation}
\tau _{22}(r_{+})=\frac{i}{\omega }\frac{4r_{+}^{4}+\mathcal{J}q^{2}}{%
4r_{+}^{4}}\tau _{32}(r_{+}).
\end{equation}%
Using the bulk action (\ref{bulkNCAC}), the Gibbons-Hawking term (\ref{GHEMA}%
) and the counter term (\ref{ctEMA}), we calculate the one-point functions%
\footnote{%
We have not taken into account any additional counterterms (if existed) due
to the non-minimal coupling. This is reasonable since the DC conductivities
have already been finite \cite{LWJ1612}.}%
\begin{eqnarray}
\frac{\delta S_{\mathrm{os}}}{\delta a_{x}} &=&-\frac{4r^{2}-\mathcal{J}%
\beta ^{2}}{4r^{2}}\left( qh_{tx}+ha_{x}^{\prime }\right) ,  \notag \\
\frac{\delta S_{\mathrm{os}}}{\delta h_{tx}} &=&\bar{C}%
_{22}h_{tx}+r^{4}h_{tx}^{\prime },  \label{onepoint3}
\end{eqnarray}%
where $\bar{C}_{22}$ is same to Eq. (\ref{C22bEMA}). Substituting Eq. (\ref%
{tau3}) into Eq. (\ref{onepoint3}), one can extract%
\begin{eqnarray}
G_{11} &=&i\omega \tau _{11},\;G_{12}=i\omega \tau _{12}-q\frac{4r^{2}-%
\mathcal{J}\beta ^{2}}{4r^{2}},  \notag \\
G_{21} &=&i\omega \tau _{21},\;G_{22}=\bar{C}_{22}+i\omega \tau _{22}.
\label{GtauNCAC}
\end{eqnarray}%
Inserting them into Eq. (\ref{Ansak}) and Eq. (\ref{OC}), we can deduce $%
\left( \hat{\sigma},\hat{\alpha},\hat{\kappa}\right) $ and $\hat{\sigma}_{%
\mathrm{0}}$ from $\tau $. Their RG flow is depicted in Figure \ref{OCNM}.

\section{Analytical DC thermal conductivity}

We have exhibited a trivial RG flow $\hat{\sigma}_{0}\left( r\right) =\hat{%
\sigma}_{0}\left( r_{+}\right) $. As a result, the DC thermal conductivity
on the boundary can be expressed as
\begin{equation}
\bar{\kappa}=\frac{T\alpha ^{2}}{\sigma -\hat{\sigma}_{0}\left( r_{+}\right)
}.  \label{FinalK}
\end{equation}%
We argue that this provides an analytical method to calculate $\bar{\kappa}$%
, if $\sigma $, $\alpha $, and $\hat{\sigma}_{0}\left( r_{+}\right) $ have
been obtained analytically. The first two ($\sigma $, $\alpha $) can be
derived in terms of the conserved electric current \cite{Donos1406}. With
the help of the regularity of the Riccati equation on the horizon, we can
write down the analytical expression of the latter one $\hat{\sigma}%
_{0}\left( r_{+}\right) $ for all the models which have been studied. More
simply, by observing the RG flow in above figures, one can find that the ZHC
conductivity meets the electrical conductivity on the horizon. So let's
write down the expression of $\sigma _{\mathrm{H}}\equiv \hat{\sigma}(r_{+})$
in those models. Using the infrared boundary condition (\ref{bc}) and the
relation between the electric conductivity and the auxiliary\ transport
matrix (\ref{sjG11}), we can read $\sigma _{\mathrm{H}}=1$ for the EMA
model. It is changed as $\sigma _{\mathrm{H}}=r_{+}$ for the GB gravity. The
change comes from the increase of the spacetime dimension instead of the GB
coupling. For the EMAD theory, one can see the effect from the gauge field
coupling, $\sigma _{\mathrm{H}}=Z\left( r_{+}\right) $. For the theory with
non-minimal coupling, $\sigma _{\mathrm{H}}=1-\mathcal{J}\beta ^{2}/\left(
4r_{+}^{2}\right) $. Combining the analytical expression of the electric
conductivity on the horizon and the boundary thermo-electric conductivities
that have been derived in \cite{Donos1406,Ge1411GB,LWJ1612}, e.g., $\sigma
=1+\frac{q^{2}}{r_{+}^{2}\beta ^{2}},\;\alpha =\frac{4\pi q}{\beta ^{2}},\;%
\bar{\kappa}=\frac{16\pi ^{2}r_{+}^{2}T}{\beta ^{2}}$ for the EMA model, one
can check $\bar{\kappa}=T\alpha ^{2}/\left( \sigma -\sigma _{\mathrm{H}%
}\right) $, as it should be.

\section{Conclusion}

We constructed a holographic RG flow of the thermo-electric transport in the
strongly coupled systems with momentum dissipation. The essence of the RG
flow is to reformulate the classical EOM in terms of the transport
coefficients measured by the intrinsic observers on the sliding membranes.
The reformulation involves two steps: recast the perturbation equations into
a Riccati equation of an auxiliary transport matrix $\tau $ and then
translate $\tau $ into the thermo-electric conductivities observed on the
membranes.

The RG flow is useful for the field theory on the boundary. First, it
provides a new method to calculate the AC thermo-electric conductivities.
Compared with the traditional method that solves the second-order
perturbation equations directly \cite{Sin1409}, the new method simplifies
the numerical calculation by just solving the first-order nonlinear ordinary
differential equation. Second, it can be used to derive the analytical
expression of the DC thermal conductivity, provided that in the DC limit the
RG flow of the ZHC conductivity does not run and the electric conductivity
and thermoelectric conductivity have been obtained analytically. Compared
with the well-known Donos-Gauntlett method \cite{Donos1406}, the RG flow
method does not need to construct the thermal current that could be subtle.

Besides the application to the boundary, the RG flow itself is interesting.
As we have shown, the RG flow of the ZHC conductivity in the DC limit does
not run for some holographic models at finite density. This generalizes the
well-known result of the membrane paradigm: the DC electrical conductivity
for neutral black holes has the trivial flow \cite{Liu0809}. We hope that
our result might provide some hints for understanding the universal
thermo-electric transport in various strongly correlated systems \cite%
{Hartnoll1612}. In particular, the $T$-linear resistivity in cuprate strange
metals can persist from near $T_{c}$ up to as high a temperature as
measured. The quick crossover from the microscopic chemistry to the
macroscopic strange-metal physics near the \textquotedblleft
ultraviolet\textquotedblright\ temperature indicates one decimation along
the RG flow in essence \cite{Zaanen1409}\footnote{%
We thank Prof. Jan Zaanen for clarifying this point to us.}.

In the future, we would like to explore whether or not the trivial RG flow
is universal when the holographic model is inhomogeneous and anisotropic.

\section*{Acknowledgments}

We thank Yi Ling, Xiao-Ning Wu, Zhuoyu Xian, and Jan Zaanen for helpful
discussions. We were supported partially by NSFC grants (No.11675097,
No.11575109, No.11375110, No.11475179, No. 11675015). YT is partially
supported by the grants (No. 14DZ2260700) from Shanghai Key Laboratory of
High Temperature Superconductors. He is also partially supported by the
\textquotedblleft Strategic Priority Research Program of the Chinese Academy
of Sciences\textquotedblright , grant No. XDB23030000. \appendix

\section{Thermodynamics on the membranes}

Here we will present the observed thermodynamics on the membranes. We start
from the positioned on-shell action. Analogue to the AdS/CFT, we define the
observed grand potential by%
\begin{equation}
\hat{\Omega}=-\hat{T}S_{\mathrm{os}},  \label{GP}
\end{equation}%
where the Tolman temperature%
\begin{equation}
\hat{T}\left( r_{c}\right) =T\frac{\Lambda (r_{c})}{\sqrt{-\gamma
_{00}(r_{c})}}  \label{Tolman}
\end{equation}%
has been invoked. We write the proper spatial volume as $\hat{V}=V_{0}\sqrt{%
\lambda _{d}}$, where $V_{0}$ denotes the spatial coordinate volume and will
be set to one for convenience. The grand potential density gives the
observed pressure $\hat{p}=-\hat{\Omega}/\hat{V}$. The observed chemical
potential should be conjugate to the observed electric charge. Based on Eq. (%
\ref{JT}), it can be written as%
\begin{equation}
\hat{\mu}=A_{\hat{t}}=A_{t}(r_{c})\frac{\Lambda (r_{c})}{\sqrt{-\gamma
_{00}(r_{c})}}.  \label{mu}
\end{equation}%
Also, one can see that the observed energy density is $\hat{\epsilon}\equiv
T^{\hat{t}\hat{t}}$.

To be clear, we will apply the observed thermodynamics to the EMA model. The
application to other models should be similar. Using the action (\ref%
{bulkEMA}), (\ref{GHEMA}), and (\ref{ctEMA}), we have found%
\begin{equation}
\hat{p}=\Lambda ^{3}\left[ \frac{h^{\prime }}{\sqrt{h}}+\frac{2\sqrt{h}}{%
r_{c}}-4+\frac{\beta ^{2}}{r_{c}^{2}}\left( 1-\frac{r_{c}-r_{h}}{\sqrt{h}}%
\right) \right] .
\end{equation}%
Using Eq. (\ref{Tolman}) and Eq. (\ref{mu}), one can express $\hat{p}$ as
the function of $\hat{T}$ and $\hat{\mu}$. This can further induce%
\begin{equation}
\left( \partial _{\hat{T}}\hat{p}\right) _{\hat{\mu}}=\frac{\Lambda ^{2}}{%
r_{c}^{2}}s,\;\left( \partial _{\hat{\mu}}\hat{p}\right) _{\hat{T}}=\frac{%
\Lambda ^{2}}{r_{c}^{2}}q,
\end{equation}%
where $s=4\pi r_{+}^{2}$. Keeping in mind $\sqrt{\lambda _{d}}%
=r_{c}^{2}/\Lambda ^{2}$ in the present, we can obtain the expected relation
for the observed thermodynamics:%
\begin{equation}
\left( \partial _{\hat{T}}\hat{p}\right) _{\hat{\mu}}=\hat{s},\;\left(
\partial _{\hat{\mu}}\hat{p}\right) _{\hat{T}}=\hat{q},
\end{equation}%
where $\hat{s}=s/\sqrt{\lambda _{d}}$ and $\hat{q}=q/\sqrt{\lambda _{d}}$
are the observed entropy density and charge density, respectively. In
particular, it implies that the total entropy $S\equiv \hat{s}\hat{V}$ is
conserved along the flow. This result recovers the assumption (the radial
variation is isentropic) proposed in Ref. \cite{Strominger1006}.
Furthermore, by calculating the observed energy density%
\begin{equation}
\hat{\epsilon}=\Lambda ^{3}\left( 4-4\frac{\sqrt{h}}{r_{c}}-\frac{\beta ^{2}%
}{r_{c}^{2}}\right)
\end{equation}%
and collecting all the observed thermodynamic quantities above, one can also
establish the Euler relation%
\begin{equation}
\hat{\epsilon}+\hat{p}=\hat{T}\hat{s}+\hat{\mu}\hat{q}.
\end{equation}

Note that the consistency of the observed thermodynamics does not depend on
the choice of the conformal factor $\Lambda ^{2}$.

\section{Semi-analytical proof}

Here we will verify semi-analytically
\begin{equation}
\partial _{r}\hat{\sigma}_{\mathrm{0}}=\mathcal{O}\left( \omega \right) .
\label{ds0A}
\end{equation}%
It is based on an assumption: up to the pole from the contact term $C_{22}$,
the canonical response functions $\Gamma _{IJ}\equiv G_{IJ}/\left( i\omega
\right) $ are finite in the DC limit. The assumption can be justified using
the numerical method.

Let's illustrate Eq. (\ref{ds0A}) in the simplest EMA model. We have to
translate the non-canonical response functions $\tau $ into the canonical
response functions $\Gamma $. However, it is difficult to inverse Eq. (\ref%
{GTau}) since it is nonlinear. Therefore, we adopt Eq. (\ref{GtauNCAC}) with
$\mathcal{J}=0$ by which $\tau $ can be represented by $G$ readily. Then the
canonical response functions can be read from $\Gamma =\frac{1}{i\omega }G$.
To subtract the pole in $\Gamma $, we define%
\begin{equation}
\tilde{\Gamma}_{22}=\frac{1}{i\omega }\left( G_{22}-C_{22}\right) ,
\label{TauC}
\end{equation}%
which is finite at $\omega \rightarrow 0$ and will be used to replace $%
\Gamma _{22}$ in the\ calculation below.

\begin{figure}[th]
\centerline{
\includegraphics[width=1\textwidth]{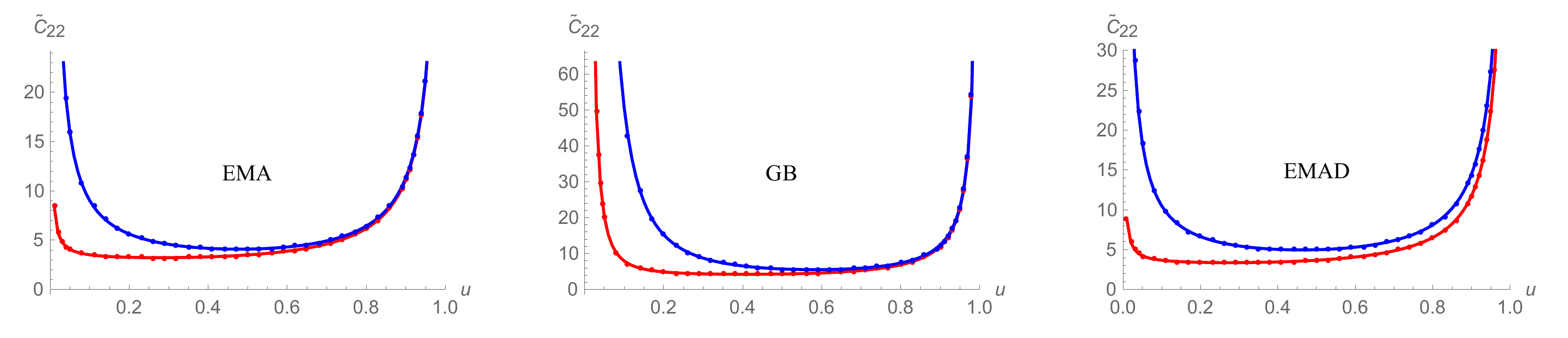}}
\caption{$\tilde{C}_{22}$ as the functions of $u=r_{+}/r$ at $\protect\omega %
/\protect\mu =10^{-4}$ for three models. The GB coupling is fixed as $\tilde{%
\protect\alpha}=9/100$. The curves denote the numerical functions $\tilde{C}%
_{22}=\lim_{\protect\omega \rightarrow 0}G_{22}-\bar{C}_{22}$ and the points
denote the analytical expressions.}
\label{C22}
\end{figure}

Putting Eqs. (\ref{Ansak}), (\ref{RiccatiNCAC}), (\ref{GtauNCAC}), and (\ref%
{TauC}) together, we can derive%
\begin{equation}
\partial _{r}\hat{\sigma}_{\mathrm{0}}=\frac{F_{1}(\Gamma )}{i\omega }\left[
r^{4}\left( q^{2}+\beta ^{2}r^{2}-h\tilde{C}_{22}^{\prime }\right) -\tilde{C}%
_{22}^{2}h\right] +F_{2}(\Gamma )\left[ \left( 2r^{3}+\tilde{C}_{22}\right)
h-r^{4}h^{\prime }\right] +\mathcal{O}(\omega ),  \label{ds0}
\end{equation}%
where $\tilde{C}_{22}\equiv C_{22}-\bar{C}_{22}$ denotes the rest of the
contact term, and%
\begin{eqnarray}
F_{1}(\Gamma ) &=&-\frac{\left( r^{2}A_{t}\Gamma _{11}-\Gamma _{12}h\right)
^{2}}{r^{4}\left[ r^{4}A_{t}^{2}\Gamma _{11}-r^{2}A_{t}\left( \Gamma
_{12}+\Gamma _{21}\right) h+\tilde{\Gamma}_{22}h^{2}\right] ^{2}}, \\
F_{2}(\Gamma ) &=&\frac{A_{t}\left( r^{2}A_{t}\Gamma _{11}-\Gamma
_{12}h\right) \left[ r^{2}A_{t}\Gamma _{11}\left( \Gamma _{12}-\Gamma
_{21}\right) -\Gamma _{12}\left( \Gamma _{12}+\Gamma _{21}\right) h-2\Gamma
_{11}\tilde{\Gamma}_{22}\right] }{r^{2}\left[ r^{4}A_{t}^{2}\Gamma
_{11}-r^{2}A_{t}\left( \Gamma _{12}+\Gamma _{21}\right) h+\tilde{\Gamma}%
_{22}h^{2}\right] ^{2}}.  \notag
\end{eqnarray}%
Using the EOM for background fields $h$ and $A_{t}$, one can find that both
terms in Eq. (\ref{ds0}) vanish if%
\begin{equation}
C_{22}=\bar{C}_{22}+r^{3}\left( \frac{rh^{\prime }}{h}-2\right) .
\label{C22EMA}
\end{equation}%
In the left panel of Figure \ref{C22}, we have checked numerically that Eq. (%
\ref{C22EMA}) is the contact term $G_{22}\left( 0\right) $ indeed. Thus, we
have demonstrated Eq. (\ref{ds0A}) by a semi-analytical method. As a bonus,
we have obtained the analytical expression of the contact term. To be more
clear, we input Eq. (\ref{C22bEMA}) into Eq. (\ref{C22EMA}). Then the
contact term is%
\begin{equation}
C_{22}=2r^{3}\left( 1-\frac{r}{\sqrt{h}}+\frac{rh^{\prime }}{2h}\right) .
\label{C22EMA2}
\end{equation}%
On the boundary, one can find%
\begin{equation}
C_{22}\left( r\rightarrow \infty \right) =\epsilon /2,  \label{c22e}
\end{equation}%
where $\epsilon $ is the energy density. The relation (\ref{c22e}) has been
obtained previously using the conserved current and the sources that are
linear in time \cite{Donos1406}. Alternatively, it can be derived in terms
of Ward identities \cite{Herzog0904,Kim1604}, if other correlators have been
known. This result can be taken as a self-consistent check of our theory.

The semi-analytical method also works for other theories. To avoid the
repetition, we neglect the details of the derivation but only give the
results. For the GB gravity, the contact term is%
\begin{equation}
C_{22}=\bar{C}_{22}+r^{2}\sqrt{\frac{f}{h}}\left( r^{2}-4\alpha f\right)
\left( \frac{rh^{\prime }}{h}-2\right) .
\end{equation}%
For the EMAD theory with the metric ansatz (\ref{EMADmetric}), the contact
term can be written as
\begin{equation}
C_{22}=\bar{C}_{22}+r^{4}f^{2}\frac{h^{\prime }}{h}.
\end{equation}%
They are both consistent with the numerical results, see the middle and
right panels in Figure \ref{C22}. On the boundary, one can check that they
are equal to $\epsilon /3$ and $\epsilon /2$, respectively. At last, note
that the non-minimal coupling does not change the contact term (\ref{C22EMA2}%
).

\end{document}